\documentclass[aps,prb,reprint,amsmath,amssymb,superscriptaddress]{revtex4-1}

\usepackage{graphicx}
\usepackage{bm}

\begin{document}


\title{Microscopic description of stress- and temperature-dependent shear modulus in solid $^4$He}


\author{Evan S. H. Kang}
\thanks{Both authors contributed equally to this work.}
\affiliation{Center for Supersolid and Quantum Matter Research and Department of Physics, KAIST, Daejeon 305-701, Republic of Korea}

\author{Hongkee Yoon}
\thanks{Both authors contributed equally to this work.}
\affiliation{Department of Physics, KAIST, Daejeon 305-701, Republic of Korea}

\author{Eunseong Kim}
\affiliation{Center for Supersolid and Quantum Matter Research and Department of Physics, KAIST, Daejeon 305-701, Republic of Korea}


\date{\today}

\begin{abstract}
We developed a detailed microscopic method that describes the shear modulus anomaly of solid helium at low temperature. The shear modulus was calculated using the pinning length of dislocations determined in detail for both crossing network nodes and $^3$He impurities. The strong suppression of the shear modulus is reproduced well as the temperature or stress increases. The shear modulus at low temperatures depends strongly on how the state was prepared. All the key features in the stress hysteresis can be quantitatively explained in terms of the thermomechanical path-dependent pinning length of dislocation networks.
\end{abstract}


\maketitle

\section{INTRODUCTION}

The observation of a shear modulus anomaly in solid $^4$He below 200 mK has attracted attention because of its remarkable similarities in temperature, $^3$He concentration, and drive dependence to the torsional oscillator response, which was previously interpreted as the appearance of non-classical rotational inertia (NCRI).\cite{1,2,3} The shear modulus anomaly can provide an alternative explanation within the classical framework for the resonant period drop without invoking superfluidity.

Despite the remarkable similarities,\cite{4,5,6} the fundamental relationship between the shear modulus and the NCRI is not clear. For instance, no evidence of a period drop was detected, although shear stiffening was also observed, in hexagonal closed packed $^3$He at low temperatures.\cite{7} Namely, decoupling can be observed only in a Bose solid, whereas stiffening occurs regardless of the quantum properties of solid helium. In addition, the response of the NCRI and shear modulus anomaly to DC rotation appeared to be very different. When DC rotation was superposed on top of AC oscillation, NCRI was dramatically suppressed, whereas the shear modulus was unaffected, indicating that the shear stiffening has a different microscopic origin from that of NCRI.\cite{8}

The characteristic features of the shear modulus anomaly can be explained by simple classical mechanisms without including superfluidity. The elastic properties of solids were originally explained by the motion of dislocations.\cite{9} Dislocations, common line defects in a solid, are the source of slippage, which softens a crystalline solid. In case of hcp helium, the dislocations are easily movable in the basal plane (0001) of the hexagonal structure.\cite{c44,c44_2} This dislocation motion leads to the reduction of the elastic coefficient $c_{44}$ and consequently affects the temperature-dependent shear modulus. The stiffening of a solid at low temperatures can be attributed to the suppression of the motion of dislocations, which can occur when the dislocation lines are pinned by impurities in the solid. These impurities can be detached by large strains or thermal fluctuations, which soften the solid.

When solid helium is not rigid but viscoelastic, its response to external AC agitation can differ from that of a rigid solid. The response from a viscoelastic solid is characterized by the relaxation model with Debye susceptibility, in which the shear modulus and dissipation are obtained as the real and imaginary parts of the susceptibility, respectively.\cite{10,11} The characteristic relaxation time grows at low temperature and causes a crossover from the relaxed to the unrelaxed phase. The drive dependence and hysteresis can then be understood in the context of a relaxation response. However, the expected change due to dissipation in solid helium is too small to explain the large shift in the shear modulus at low temperatures.\cite{10} A more complicated type of susceptibility, such as glassy susceptibility with suitable extra parameters, is necessary to understand the inconsistency between the quantitative magnitude of the shear modulus and the dissipation.\cite{12,13} The shear modulus and the dissipation with a relatively reasonable ratio were recently predicted by applying a distribution of the binding energy between $^3$He impurities and dislocations.\cite{6}

Iwasa, on the other hand, suggested a simplified model to describe the temperature- and $^3$He-concentration-dependent shear modulus on the basis of the Granato--Lucke (GL) model.\cite{9,14} In this vibrating dislocation model, the vibration of dislocations is essentially characterized by the average dislocation length, which was approximated by the average network pinning length at high temperatures and by the average impurity pinning length at low temperatures. The crossover is induced by progressive shortening or lengthening of the average pinning length. This mostly gradual change in the total fraction of $^3$He impurities binding on the dislocations is determined by the temperature and $^3$He concentration. The shorter pinning length at low temperatures restricts the motion of dislocations and, accordingly, stiffens the solid helium. The dissipation, which is equivalent to the damping of the dislocation motion in this model, shows good agreement with the observed experimental results.

\begin{figure*}[t]
\includegraphics[width=16cm]{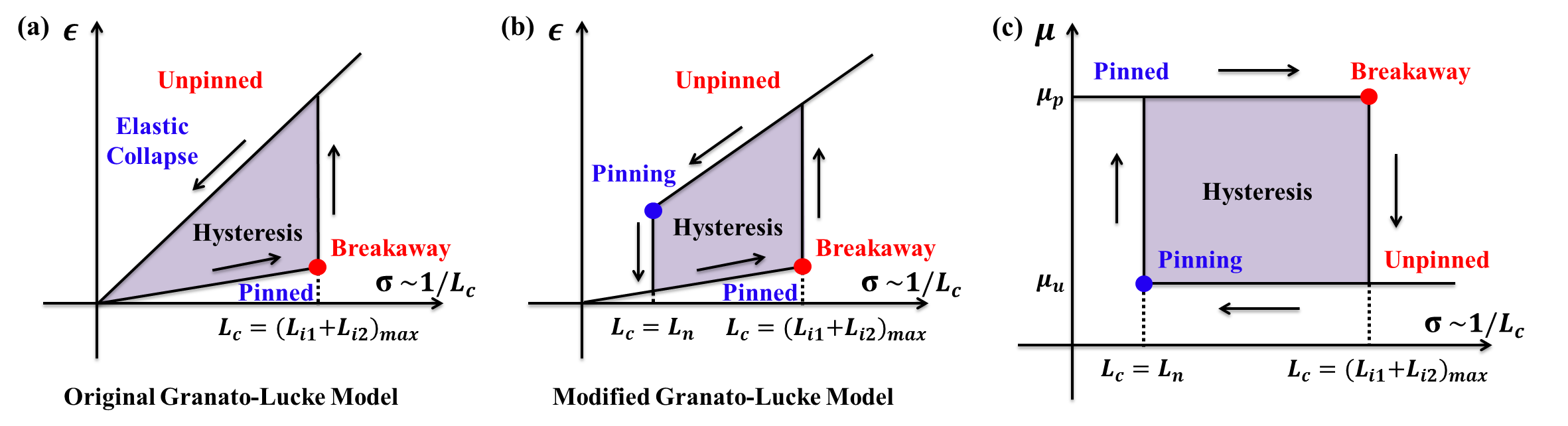}
\caption{\label{fig:1} Schematic stress--strain relation for a single network line using (a) the original Granato--Lucke model and (b) the modified model. (c) Shear modulus for a single network line calculated using the modified model.}
\end{figure*}

Although all the models mentioned above qualitatively explain the characteristics of the shear modulus, the unusual drive- and temperature-dependent hysteresis is difficult to understand with these simplified models. Here, we investigated the microscopic description of the drive- and temperature-dependent hysteresis on the basis of dislocation pinning by $^3$He impurities. Experimental results, including the hysteresis, drive dependence, temperature dependence of the shear modulus, and dissipation, are reproduced entirely with a single set of fitting parameters.

\section{THE MODEL}
\subsection{Original Granato--Lucke model\cite{9}}

Dislocations in a solid constitute a complicated three-dimensional network. The motion of dislocations is strongly pinned in the nodes (the intersections of the network). The dislocations can be more weakly pinned by impurities, which depends strongly on the temperature and stress in the solid. Accordingly, the dynamics of the dislocations can alter the elastic properties of a solid significantly. The external shear stress $\sigma$ applied to a solid induces a strain that can be represented by the sum of the elastic strain and an additional strain due to the vibrating dislocation loops,
\begin{equation} \label{eq:1}
\epsilon_{total} = \epsilon_{el}+\epsilon_{dis}.
\end{equation}
Because the contribution of a single dislocation loop of length $L$ to the strain is proportional to $L^{3}$, the total additional strain from dislocations is simply expressed as the sum of each contribution,\cite{9}
\begin{equation} \label{eq:2}
\epsilon_{dis} = \sum CRL^{3} \frac{\sigma}{\mu_{0}},
\end{equation}
where $C=\frac{4(1-\nu)}{\pi^{3}}$, $\nu$ is Poisson$'$s ratio (= 0.3), $R$ is the orientation factor, and $\mu_{0}$ is the elastic shear modulus. For the ramdom orientations in polycrystalline solids, the anisotropic effect can be averaged into $R$ $(\sim 0.2)$.\cite{4} This dislocation-induced strain decreases the shear modulus compared to that in the perfect crystalline solid with no dislocation. Using Eqs. (1) and (2), one may write the normalized shear modulus (the shear modulus $\mu$ divided by the elastic shear modulus $\mu_{0}$) in terms of the loop length,
\begin{equation} \label{eq:3}
\frac{\mu}{\mu_{0}} = \Big[1+\sum CRL^{3}\Big]^{-1}.
\end{equation}
In the absence of impurity atoms, the characteristic dislocation loop length is determined solely by the length $L_{n}$ of the dislocation network link between two network nodes. When a sufficient number of impurity atoms, such as $^3$He atoms, are present, they diffuse into the strained region near the dislocation lines and pin the dislocation lines, functioning as pinning centers. Consequently, the average length of the dislocation loops is reduced and gradually approximates the impurity pinning length $L_{i}$ determined by the loop length between two impurity pinning sites. Because the additional strain is directly related to the length of the dislocation loops, the pinning of dislocations by impurities will suppress the total strain significantly. Accordingly, the shear modulus of a solid with fully pinned dislocations is nearly the same as that of a perfect crystal.

Given that the external shear stress on a pinning point is greater than the binding force between a dislocation loop and an impurity, the stress can break the dislocation away from the impurity pinning point. Because the tension on a pinning point induced by shear stress is proportional to the total length of two neighboring loops, $L_{i1} + L_{i2}$, dislocations shorter than a characteristic critical length $L_{c}$ remain pinned. The critical length is determined by the maximum binding force $f_{m}$, and external stress $\sigma$, given by $L_{c}(\sigma)=\pi f_{m}/4b\sigma$, where $b$ is the magnitude of Burger’s vector. The breakaway condition in terms of the critical length is, therefore,
\begin{equation} \label{eq:4}
L_{i1}+L_{i2}>L_{c}.
\end{equation}
Granato and Lucke pointed out that this stress-induced breakaway is a catastrophic process. For instance, once the breakaway condition is satisfied, a dislocation will be detached from the corresponding pinning site and produce a longer loop $(L_{i1}+L_{i2})$. The resultant loop will constitute a new longer pair with another adjacent loop $(L_{i1}+L_{i2}, L_{i3})$ that obviously satisfies the breakaway condition, $L_{i1} + L_{i2} + L_{i3} > L_{c}$. This chain-reaction process continues catastrophically until the entire impurity-pinned dislocation is unpinned. Because any pair of adjacent loops that comply with the breakaway criterion can trigger this catastrophic development, the sufficient condition for this transition in a dislocation line is given by
\begin{equation} \label{eq:5}
(L_{i1}+L_{i2})_{max}>L_{c},
\end{equation}
where the subscript $max$ indicates the maximum resultant length among all the adjoining pairs in the dislocation line.

On the other hand, a gradual decrease in stress causes progressive pinning of dislocations. The stress--strain relation for a single network link in the GL model is delineated in Fig. \ref{fig:1}(a).

\subsection{Modified Granato--Lucke model for solid helium}

The GL model is a pioneering model explaining the macroscopic physical quantities using the microscopic behavior of dislocations. Nevertheless, minor modification is unavoidable when the model is applied to the solid helium system. As shown in Fig. \ref{fig:1}(a), dislocations are not pinned until the applied stress decreases back to zero in the original GL model. This is because the point defects in the solid are not mobile. We assume that $^3$He atoms, the prevailing impurities in solid $^4$He, can approach dislocations via quantum diffusion, unlike impurities in a normal solid.\cite{15} This peculiar property allows pinning of dislocations even under non-zero stress, as shown in Fig. \ref{fig:1}(b). The pinning rate can be determined by the competition between the binding energy of impurities to dislocations and the vibrational energy that tears impurities from dislocations. Suppose a $^3$He atom sticks to a dislocation link that is pinned to two neighboring network nodes only. The tension, which is greater than the binding force, will remove the $^3$He impurity from the dislocation link immediately after binding. The dislocation link can remain pinned only if the exerted tension is sufficiently small. As described in the unpinning process of the original GL model, the tension is proportional to the total length of two adjoining impurity pinning loops, which in this case is the length of the network link. Hence, the pinning condition can be proposed in terms of the critical length as well:
\begin{equation} \label{eq:6}
L_{n}<L_{c}.
\end{equation}

Once a dislocation network link is pinned by a $^3$He atom, the link will be separated into two shorter loops, $L_{i1}$ and $L_{i2}$, both of which again satisfy the pinning condition, $L_{i1}, L_{i2} < L_{c}$. Therefore, the pinning process in solid helium is also catastrophic. For easy comparison, the stress--strain relation for a single network line using the modified GL model is delineated in Fig. \ref{fig:1}(b).

In Fig. \ref{fig:1}(c), the shear modulus $\mu$ is obtained as a function of the stress by simply using the relation, $\mu=\sigma /\epsilon$, where $\sigma$ is the external stress, and $\epsilon$ is the total induced strain. Before the breakaway point in a stress up-scan, a single dislocation link shows a relatively high, stress-independent shear modulus $(\mu_{p})$. As the stress reaches the breakaway condition, the dislocation link is softened by the unpinning of the $^3$He impurities, with a lower shear modulus $(\mu_{u})$. The abrupt change of the shear modulus in a stress down-scan can be explained in essentially the same way. As the stress decreases below the pinning point, the dislocation is stiffened again owing to the reduced loop lengths and recovers the previous value of the shear modulus $(\mu_{p})$. Note that the hysteretic behavior of the shear modulus described in Fig. \ref{fig:1}(c) is not completely new but rather is similar to the hysteretic elastic unit (HEU) in the Preisach--Mayergoyz space model.\cite{16,17} In this model, the entire system is assumed to consist of a large number of HEUs, and its macroscopic elastic properties are considered to be the integral response of individual HEUs.\cite{18}

\begin{figure}[b]
\includegraphics[width=9cm]{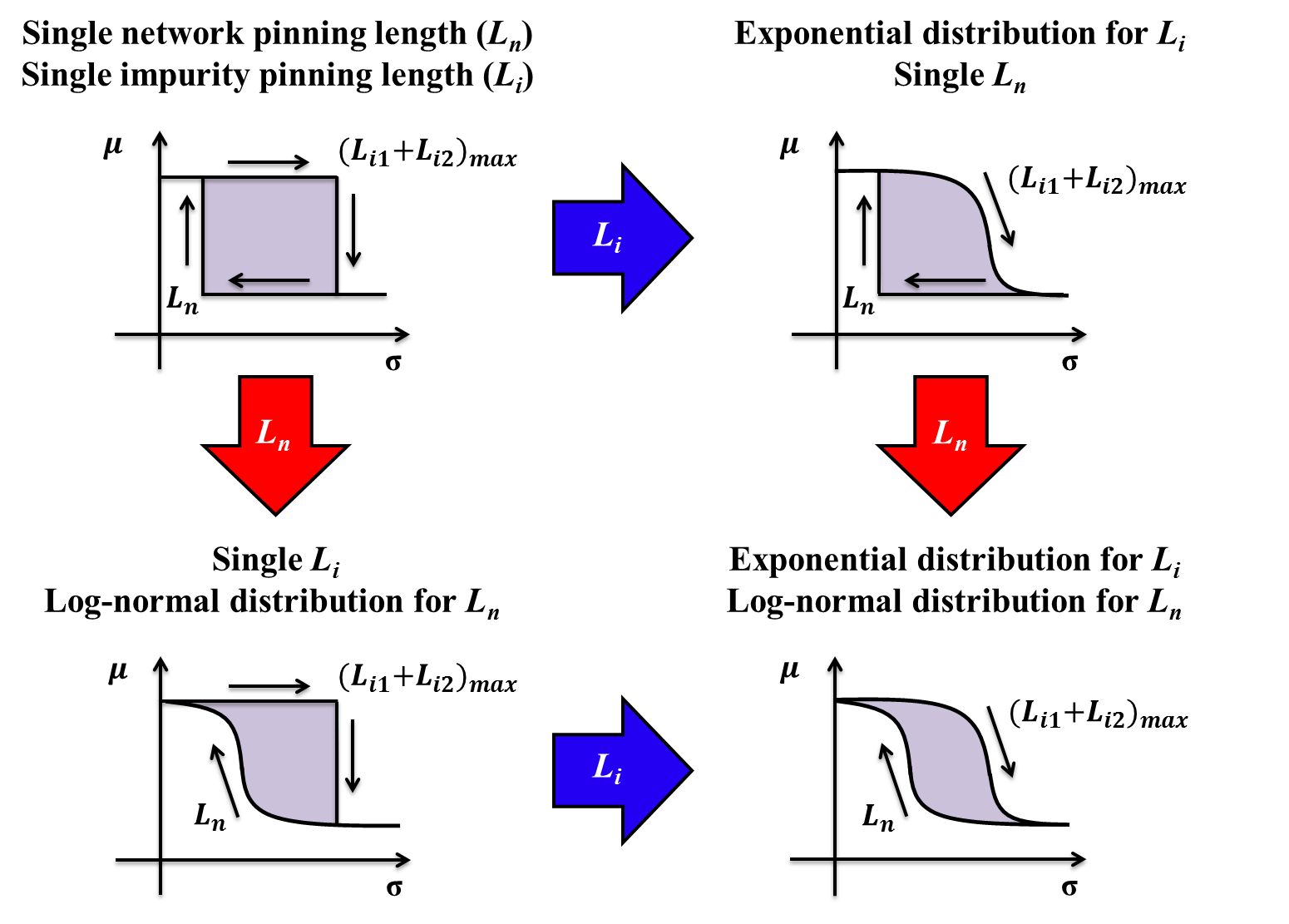}
\caption{\label{fig:2} Distribution effect on the shear modulus as a function of stress.}
\end{figure}

\section{DISTRIBUTION OF DISLOCATION PINNING LENGTH}

The hysteresis of the shear modulus shown in Fig. \ref{fig:1}(c) is limited to a model with a single dislocation link. A more realistic description should consider the distributions of both impurity pinning lengths $L_{i}$ and network pinning lengths $L_{n}$. The loop length distribution dramatically changes the shape of the hysteresis, as shown in Fig. \ref{fig:2}.

For randomly arranged solvent and impurity atoms along a one-dimensional dislocation line, the distribution of the length between the impurities (which is equivalent to $L_{i}$) follows an exponential function.\cite{19} The distribution function of the impurity pinning loops is written as $N(L_{i})dL_{i}=\frac{\Lambda}{L_{iA}^2}\exp⁡(-\frac{L_{i}}{L_{iA}})dL_{i}$, where $\Lambda$ is the total length of dislocations per unit volume, i.e., the dislocation density, and $L_{iA}$ is the average length of $L_{i}$. When network pinning is superposed on impurity pinning, the pre-existing impurity pinning loops are further separated into shorter loops by network nodes, resulting in a more complicated distribution rather than the simple exponential form. Only the average value of the pinning length $L=\frac{L_{n}L_{i}}{L_{n}+L_{i}}$ has hitherto been exploited to calculate the temperature- and impurity-concentration-dependent shear modulus because of this complexity.\cite{14}

This oversimplified model cannot accurately describe the real solid system, as illustrated in Fig. \ref{fig:2}. For instance, the smoothed transition from the pinned to the unpinned state with increasing applied stress does not appear in this model. Thus, we developed a more realistic model that allows us to consider various network and impurity pinning lengths of dislocation lines.

First, all dislocation segments between nodes in the network can be detached and then connected to reconstruct a one-dimensional dislocation line with length $\Lambda$. All the network nodes can be superposed on this one-dimensional dislocation line. The links between the nodes contain the original length distribution of a specific dislocation length $L_{n}$. When the dislocation links are arranged in ascending order of length, the links with similar length can be categorized in a group. We will consider a specific group of dislocation links with a network pinning length $L_{n}$ initially. Later, the discussion will be extended to the entire dislocation network to deal with a more realistic model.

\begin{figure}[t]
\includegraphics[width=8cm]{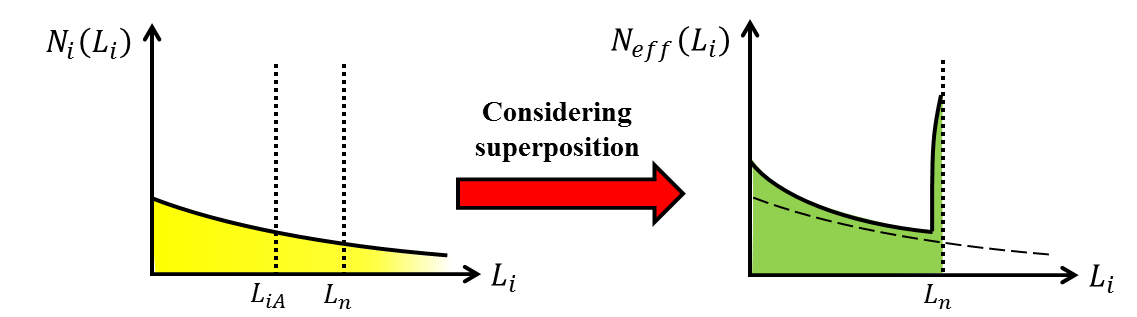}
\caption{\label{fig:3} Effective distribution function due to the superposition of the network pinning points on the impurity pinning length.}
\end{figure}

Next, additional pinning of dislocation links by impurities should be considered. The impurity pinning length $L_{i}$ is defined by the dislocation segments pinned by two neighboring impurities. For a given specific group of dislocation links with a network pinning length $L_{n}$, we find that the effective pinning length cannot exceed $L_{n}$ because the impurity pinning length $L_{i}$ longer than $L_{n}$ will be split by network nodes. Specifically, it is divided into several segments of dislocation with length $L_{n}$ and segments shorter than $L_{n}$. In addition, $L_{i}$ shorter than $L_{n}$ can also be divided by the presence of network nodes when a set of impurity distributions is superposed with the specific network pinning length $L_{n}$. (See Appendix.) Accordingly, it is necessary to define a new effective distribution function for the impurity pinning length in a specific dislocation network link, as shown in Fig. \ref{fig:3}.

The new impurity pinning length distribution can be obtained by eliminating all the impurity pinning lengths split by network nodes from the original distribution and adding newly generated dislocation pinning lengths. (See Appendix for a detailed derivation.) The new distribution function can be written as
\begin{widetext}
\begin{equation}\label{eq:7}
\begin{array}{ll}
N_{p}(L_{i},L_{n})dL_{i} = (1-\frac{L_{i}}{L_{n}}+\frac{2L_{iA}}{L_{n}})\frac{\Lambda_{n}}{L_{iA}^2}\exp⁡(-\frac{L_{i}}{L_{iA}})dL_{i}, & \textrm{for } 0<L_{i}<L_{n} \\\\
N_{u}(L_{n}) = \frac{\Lambda_{n}}{L_{n}}\exp⁡(-\frac{L_{n}}{L_{iA}}), & \textrm{for } L_{i}=L_{n}
\end{array}
\end{equation}
\end{widetext}
where $\Lambda_{n}dL_{i} = L_{n}N_{n}(L_{n})dL_{n}$ is the total length of the dislocations per unit volume (i.e., the dislocation density) of which the dislocation links have lengths in the range $(L_{n}, L_{n} + dL_{n})$; $N_{n}(L_{n})dL_{n}$ is the distribution function of the network links with length $L_{n}$, which will be discussed below, and $L_{iA}$ is the average value of $L_{i}$. The upper distribution function $N_{p}(L_{i}, L_{n})$ indicates the effective distribution function of impurity-pinned loops with lengths in the range $(L_{i}, L_{i} + dL_{i})$ when the dislocation network pinning length is between $L_{n}$ and $L_{n}$ + d$L_{n}$. The lower distribution function, $N_{u}(L_{n})$, represents the number of impurity-free unpinned links of length $L_{n}$ even at zero stress. Note that the distribution function satisfies the following normalization condition:
\begin{equation} \label{eq:8}
\int_{0}^{L_{n}}L_{i}N_{p}(L_{i},L_{n})dL_{i}+L_{n}N_{u}(L_{n})=\Lambda_{n}.
\end{equation}
Now we extend the discussion to the dislocation network with various dislocation network pinning lengths. For a polycrystalline solid, the length of the dislocation network links are known to follow a log-normal distribution,\cite{20}
\begin{equation} \label{eq:9}
N_{n}(L_{n})dL_{n}=Z\exp\Big[-\frac{(\ln{L_{n}}-\ln{\bar{L_{n}}})^{2}}{s^{2}}\Big]dL_{n},
\end{equation}
where $s$ is the width of the distribution, and $Z$ and $\bar{L_{n}}$ are given by $\frac{\Lambda}{\sqrt{\pi}sL_{nA}^2}\exp(\frac{s^{2}}{2})$ and $L_{nA}\exp(-\frac{3s^{2}}{4})$, respectively. $L_{nA}$ is the average value of $L_{n}$ and $\Lambda=\int_{0}^{\infty}\Lambda_{n}dL_{n}=\int_{0}^{\infty}L_{n}N_{n}(L_{n})dL_{n}$ is the total length of the dislocations per unit volume. Ultrasonic measurements have revealed that the average network pinning length $L_{nA}$ and the width $s$ were several micrometers and in the range of (0.5--1.3), respectively.\cite{20} Typical values for $\Lambda$ are $10^{9}$--$10^{10}$ m$^{-2}$ for relatively low-quality crystals.\cite{21} Combining the above distribution functions for $L_{i}$ and $L_{n}$ enables us to access the microscopic loop length configurations by performing rather simple integrations. Note that the distribution functions satisfy the following normalization condition:
\begin{widetext}
\begin{equation} \label{eq:10}
\int_{0}^{\infty}\Lambda_{n}(L_{n})dL_{n}=\int_{0}^{\infty}[\int_{0}^{L_{n}}L_{i}N_{p}(L_{i},L_{n})dL_{i}+L_{n}N_{u}(L_{n})]dL_{n}=\Lambda.
\end{equation}
\end{widetext}

\section{RESULTS AND DISCUSSION}
\subsection{Effect of temperature and stress on loop length configurations}

\begin{figure*}
\includegraphics[width=16cm]{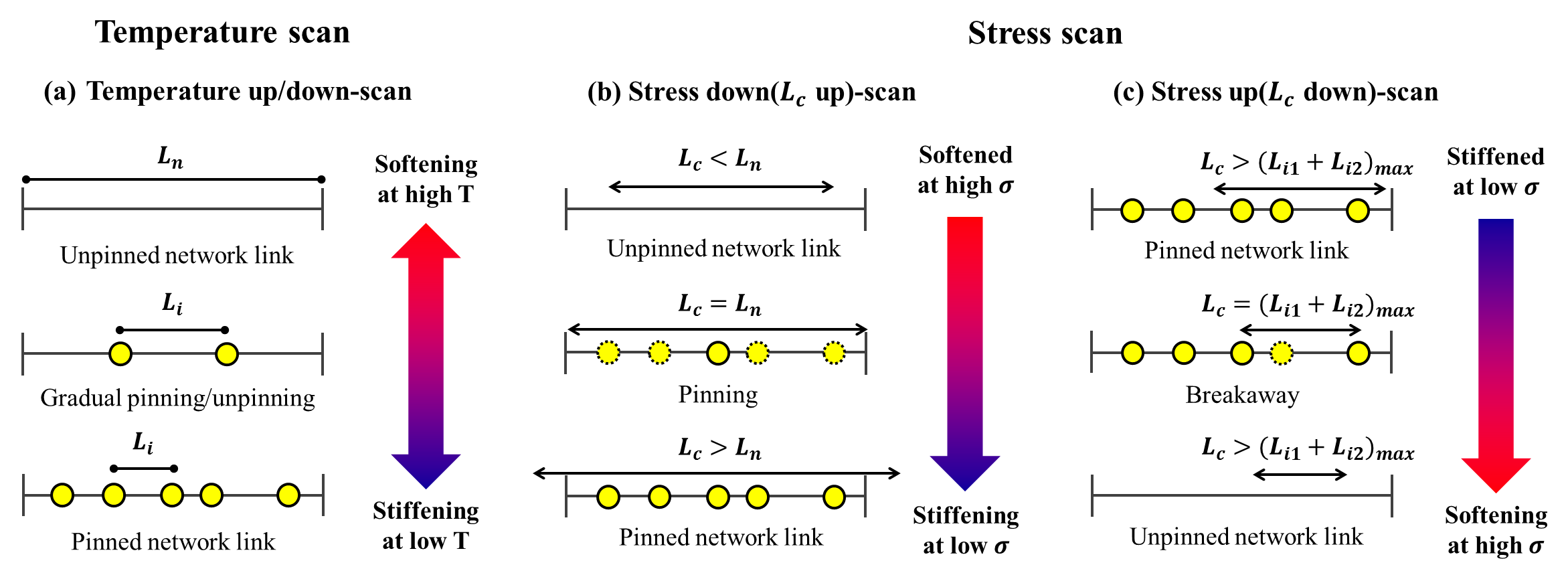}
\caption{\label{fig:4} Microscopic pinning and unpinning process on a single network link in (a) temperature scan, (b) stress up-scan, and (c) stress down-scan.}
\end{figure*}

Conventionally, two types of measurement procedures are employed for systematic study of the drive dependence and temperature dependence of the shear modulus: (1) a temperature scan and (2) a stress scan. In the temperature scan, a cell containing solid helium is cooled to the base temperature under a constant stress, and then the shear modulus is measured during warmup. In the stress scan, the cell is cooled to the desired temperature under the maximum stress, and then the shear modulus is measured during the stress down/up cycle. The detailed procedure is described elsewhere.\cite{22}

The microscopic configuration in the pinning/unpinning process on a network link during the temperature scan is depicted in Fig. \ref{fig:1}(a). At temperatures higher than approximately 500 mK, $^3$He atoms evaporate from the dislocations, and the dislocations become pinned only by the network nodes or jogs.\cite{jog} The pinning by jogs is not discussed in this manuscript for the simplicity. All the pinning loops then have the length distribution of network links, $N_{n}(L_{n})$, and the resulting normalized shear modulus becomes $[1+\sum CRL_{n}^{3}]^{-1}$. If all the network links have the same constant network length $L_{n0}$, the normalized shear modulus will be $[1+\sum CRL_{n0}^{3}]^{-1}=[1+CR\Lambda L_{n0}^{2}]^{-1}$, which leads to the previous result, $\Delta\mu/\mu = -CR\Lambda L_{n0}^{2}$.\cite{1,4} Because the network nodes function as strong pinning points, each $L_{n}$ is considered to be a constant throughout the entire experiment unless the dislocation network is changed by irreversible plastic deformation. As the temperature decreases, because the binding energy becomes dominant compared to the $^3$He kinetic energy of thermal motion, more $^3$He atoms tend to pin the dislocation lines. As a result, the average impurity pinning length $L_{iA}$ decreases depending on the temperature T according to $L_{iA}(T) = \frac{a}{x_{3}}\exp(-\frac{E_{b}}{T})$, where $a$ is the interatomic distance in solid helium, $x_{3}$ is the $^3$He concentration, and $E_{b}$ is the binding energy between a dislocation link and an impurity atom. The decrease in $L_{iA}$ significantly reduces the additional strain because the additional strain depends on the third power of the dislocation lengths, as mentioned above. Consequently, the normalized shear modulus increases to $[1+\sum CRL_{i}^{3}(T)]^{-1}$, which is much closer to one than the shear modulus at high temperature, $[1+\sum CRL_{n}^{3}]^{-1}$. Note that since the relaxation time in the $^3$He pinning/unpinning process is extremely short compared to the measurement frequencies, $L_{iA}$ varies continuously as the temperature changes. Therefore, no hysteresis appears in the temperature scan, which is consistent with experiments.\cite{23}

A similar analogy can be found in the stress-induced pinning/unpinning process, as shown in Fig. \ref{fig:4}(b). In the high-stress limit, all the links become unpinned, and the normalized shear modulus saturates at $[1+\sum CRL_{n}^{3}]^{-1}$, which is independent of temperature. In the low-stress limit, all the links become pinned, and the normalized shear modulus approaches $[1+\sum CRL_{i}^{3}(T)]^{-1}$, which decreases monotonically as the temperature increases. In the intermediate stress range, a smooth transition of the shear modulus can be found depending on the pinning length distributions. Unlike the temperature scan, the stress scan always involves the hysteresis in this regime owing to the different pinning/unpinning conditions outlined above. Hereafter, the shear modulus for each scan will be calculated quantitatively by considering the microscopic loop-length configurations.

\subsection{Pinning process: The stress down-scan}

Assume a dislocation network whose links are of various lengths following a log-normal distribution function, $N_{n}(L_{n})$. At sufficiently high stress, i.e., for low critical length, all the links will vibrate freely without any impurity pinning. As the stress decreases, shorter links satisfying the pinning condition $(L_{n}<L_{c})$ will be pinned earlier. When the system reaches the desired stress during the stress down-scan, all the links shorter than the critical length will become pinned, but not those longer than the critical length. Then the normalized shear modulus, Eq. (\ref{eq:3}), can be expressed as
\begin{equation} \label{eq:11}
\frac{\mu}{\mu_{0}}=\Big[1+CR\Big(\sum_{L_{n}<L_{c}}L^{3}+\sum_{L_{n}>L_{c}}L^{3}\Big)\Big]^{-1},
\end{equation}
where
\begin{displaymath}
\sum_{L_{n}<L_{c}}L^{3}=\int_{0}^{L_{c}}\Big[\int_{0}^{L_{n}}L_{i}^{3}N_{p}(L_{i},L_{n})dL_{i}+L_{n}^{3}N_{u}(L_{n})\Big]dL_{n},
\end{displaymath}
\begin{displaymath}
\sum_{L_{n}>L_{c}}L^{3}=\int_{L_{c}}^{\infty} L_{n}^{3} N_{n}(L_{n})dL_{n}.
\end{displaymath}

\subsection{Breakaway process: The stress up-scan}

At sufficiently low stress, all the dislocation network links will be pinned, satisfying the pinning condition. As the stress increases, adjoining loop pairs satisfying the breakaway condition $((L_{i1}+L_{i2})_{max}>L_{c})$ will trigger the breakaway of the entire link including them. The probability that a pair $(L_{i1}, L_{i2})$ stays pinned is the same as the probability that the total length of the pair, $(L_{i1} + L_{i2})$, is less than the critical length $L_{c}$:
\begin{widetext}
\begin{equation} \label{eq:12}
Prob_{p}(L_{n},L_{c})=\frac
{\int_{0}^{L_{c}}dL_{i1}N_{p}(L_{i1},L_{n})\times\int_{0}^{L_{c}-L_{i1}}dL_{i2}N_{p}(L_{i2},L_{n})}
{\int_{0}^{L_{n}}dL_{i1}N_{p}(L_{i1},L_{n})\times\int_{0}^{L_{n}}dL_{i2}N_{p}(L_{i2},L_{n})},
\textrm{ for } L_{n}>L_{c}.
\end{equation}
\end{widetext}
The length of any pair on the link cannot exceed $L_{n}$ [i.e., $(L_{i1}+L_{i2})<L_{n}$]; therefore, when $L_{n} < L_{c}$, the breakaway condition $((L_{i1}+L_{i2})_{max}>L_{c})$ cannot be satisfied, and $Prob_{p}(L_{n},L_{c})$ is equal to one. Because any adjoining pair on the link could trigger the breakaway, the probability that the breakaway is not triggered by any adjoining pairs on the link is $[Prob_{p}]^{k}$. Here $k$ is the average number of adjoining pairs on a link; it is given by $k=\frac{L_{n}}{L_{iA}}[1-\exp⁡(-\frac{L_{n}}{L_{iA}})]^{-1}$, which is the same as the average number of impurity points on a link. (See Appendix for a detailed derivation.) In other words, the $[Prob_{p}]^{k}$ fraction of the pinned links stays pinned in the breakaway process, whereas the $(1-[Prob_{p}]^{k})$ fraction becomes unpinned. The loop length distribution for the breakaway is then modified into a new form,
\begin{widetext}
\begin{equation}\label{eq:13}
\begin{array}{ll}
N_{p}'(L_{i},L_{n})dL_{i} = [Prob_{p}]^{k} \times N_{p}(L_{i},L_{n})dL_{i}, & \textrm{for } 0<L_{i}<L_{n} \\\\
N_{u}'(L_{n}) = N_{u}(L_{n}) + (1-[Prob_{p}]^{k}) \times \frac{\Lambda_{n}}{L_{n}}[1-\exp(-\frac{L_{n}}{L_{iA}})] , & \textrm{for } L_{i}=L_{n}
\end{array}
\end{equation}
\end{widetext}
where $\frac{\Lambda_{n}}{L_{n}}[1-\exp(-\frac{L_{n}}{L_{iA}})]$ is the total number of pinned links before the breakaway, which is calculated in Appendix. Note that the total length of the dislocations is conserved because the reduced length in the upper distribution is compensated by the supplemented length in the lower distribution. One may find that if $L_{n} < L_{c}$ and $[Prob_{p}]^{k}$ equals one, the loop length distribution reverts to the previous distribution form. Then the normalized shear modulus may be expressed as
\begin{equation} \label{eq:14}
\frac{\mu}{\mu_{0}}=\Big[1+CR\Big(\sum_{L_{n}<L_{c}}L^{3}+\sum_{L_{n}>L_{c}}L^{3}\Big)\Big]^{-1},
\end{equation}
where
\begin{displaymath}
\sum_{L_{n}<L_{c}}L^{3}=\int_{0}^{L_{c}}\Big[\int_{0}^{L_{n}}L_{i}^{3}N_{p}(L_{i},L_{n})dL_{i}+L_{n}^{3}N_{u}(L_{n})\Big]dL_{n},
\end{displaymath}
\begin{displaymath}
\sum_{L_{n}>L_{c}}L^{3}=\int_{L_{c}}^{\infty}\Big[\int_{0}^{L_{n}}
L_{i}^{3}N_{p}'(L_{i},L_{n})dL_{i}+L_{n}^{3}N_{u}'(L_{n})\Big]dL_{n}.
\end{displaymath}

\begin{figure}[t]
\includegraphics[width=7.5cm]{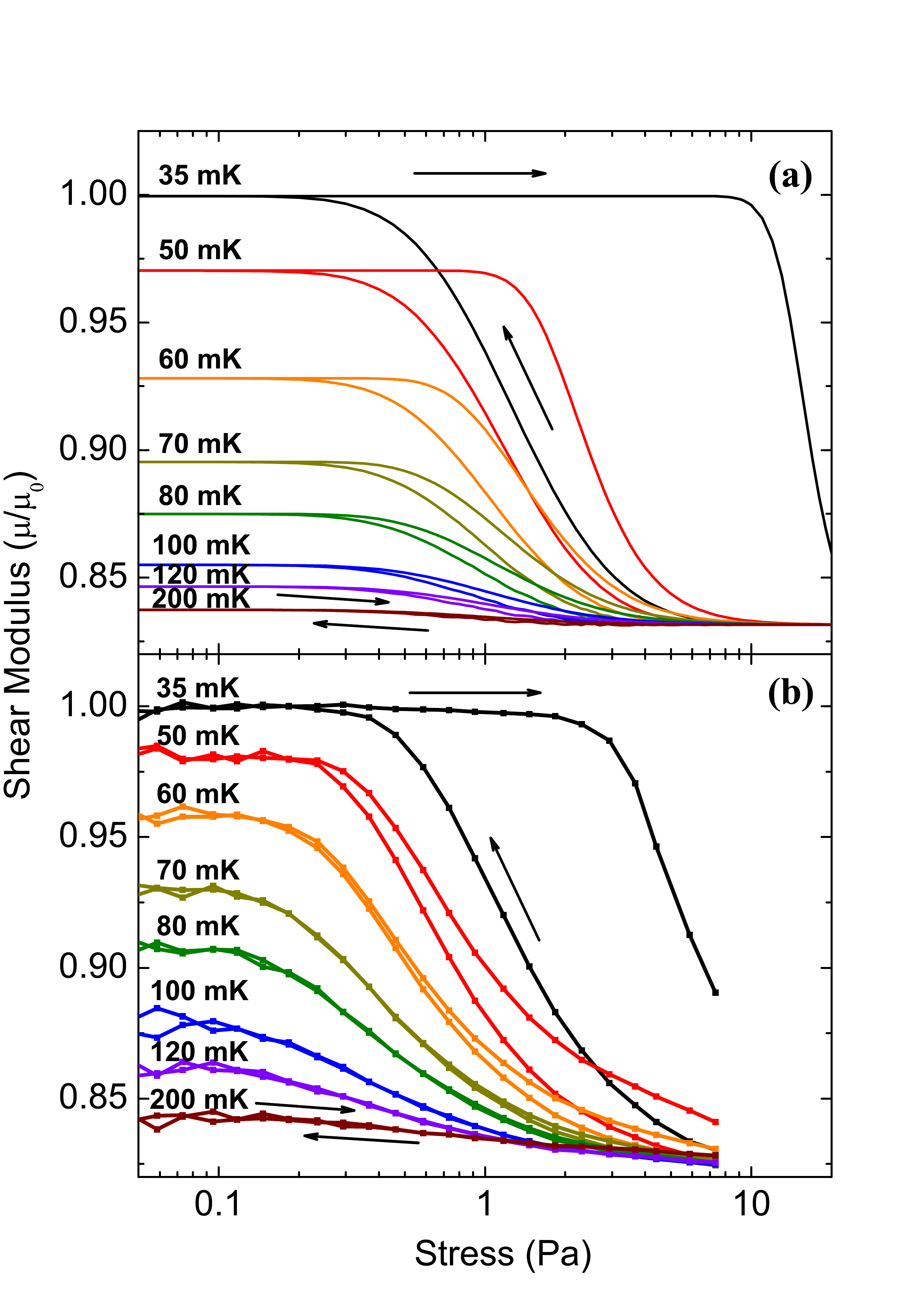}
\caption{\label{fig:5} (a) Shear modulus as a function of stress calculated using Eqs. (11) and (14). (b) Experimental shear modulus data measured during the stress down/up-scan.\cite{22}}
\end{figure}

The normalized shear modulus in the pinning/breakaway process is given as a function of $L_{iA}(T)$ and $L_{c}(\sigma)$. The calculated results are displayed in Fig. \ref{fig:5}(a) as a function of $T$ and $\sigma$ instead of $L_{iA}$ and $L_{c}$ using the relations, $L_{iA}(T)=\frac{a}{x_{3}}\exp⁡(-\frac{E_{b}}{T})$ and $L_{c}(\sigma)=\pi f_{m}/4b\sigma$, respectively. According to the classical elasticity theory proposed by Cottrell, the maximum attractive force between an impurity and a dislocation is $f_{m}=-3\sqrt{3}E_{b}/8r_{0}$, where $r_{0}$ is the closest distance between the impurity and the dislocation center, which is given by $b/\sqrt{2}$, and $E_{b}$ is the binding energy.\cite{24} For the best fit to the experimental data, the parameters in the distributions were obtained as follows: the binding energy $E_{b}=0.3$ K, the average network link length $L_{nA}=5$ $\mu$m, the width of the network link length distribution $s=1$, and $R\Lambda=2.01\times10^{10}$ m$^{-2}$. The experimental results are shown in Fig. \ref{fig:5}(b) for comparison. The arrows indicate the scan direction during each measurement. The stress range and temperature range of the crossover from the pinned state with high shear modulus to the unpinned state with low shear modulus show surprisingly good agreement with the calculated results.

\begin{figure}[b]
\includegraphics[width=7.5cm]{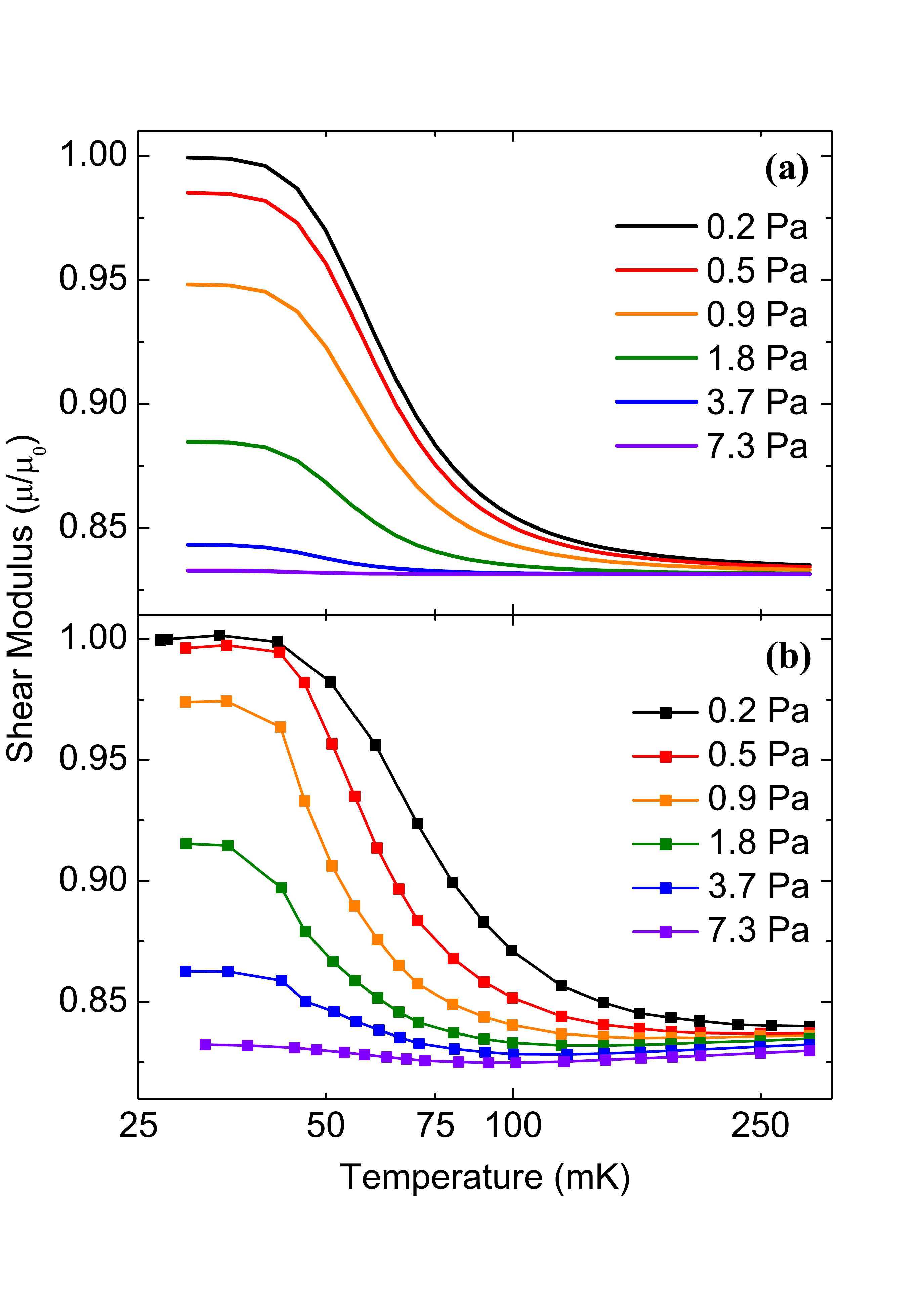}
\caption{\label{fig:6} (a) Shear modulus as a function of temperature calculated using Eq. (\ref{eq:11}). (b) Experimental shear modulus data measured during the temperature scan.\cite{22}}
\end{figure}

\begin{figure}[t]
\includegraphics[width=8cm]{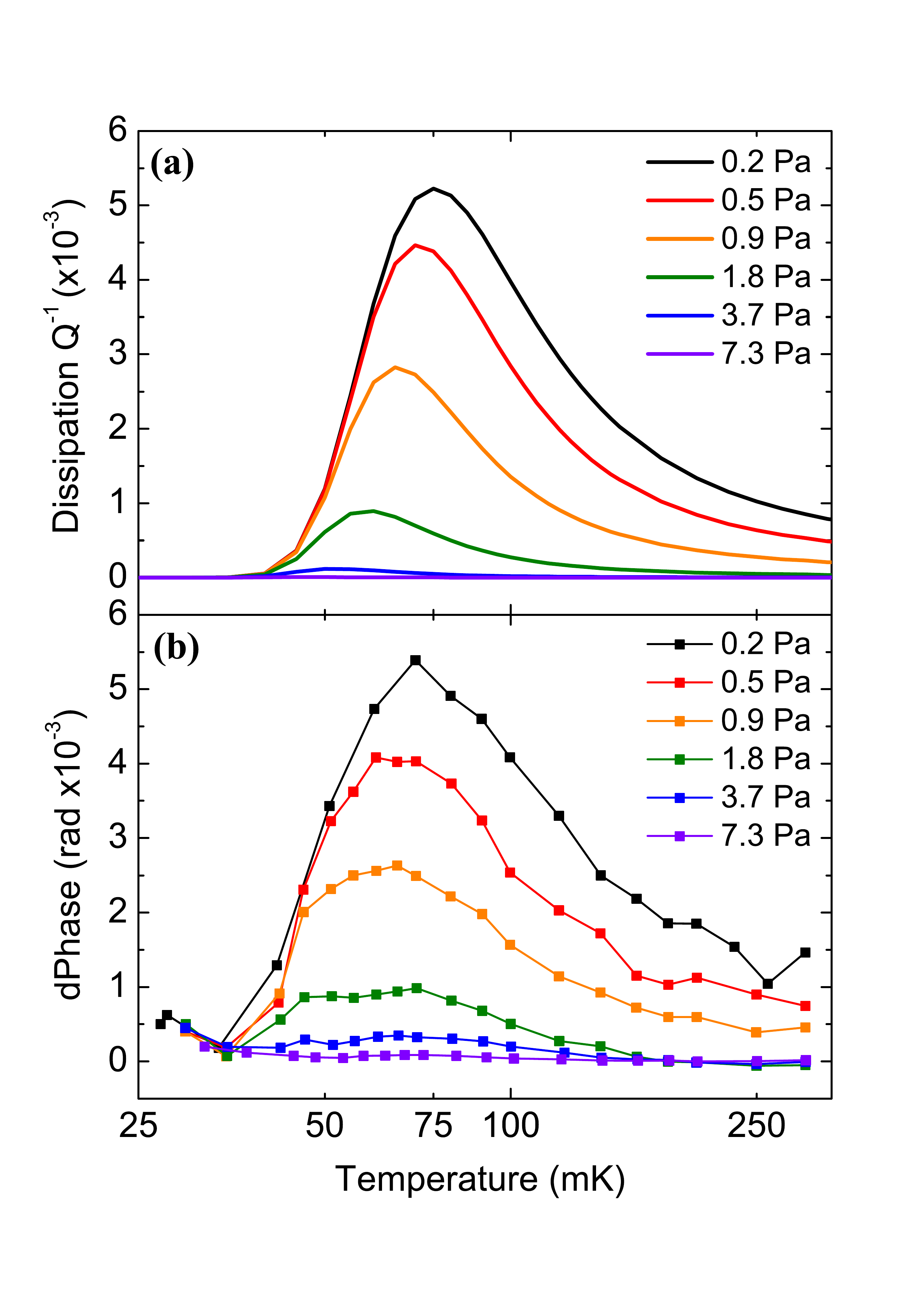}
\caption{\label{fig:7} (a) Dissipation as a function of temperature calculated using Eq. (\ref{eq:18}). (b) Experimentally measured phase shift during the temperature scan.\cite{22}}
\end{figure}

\subsection{Constant stress process: The temperature scan}

The conventional temperature scan resembles the pinning process in the mechanical breakaway model. Note that, in the temperature scan with a fixed stress [Fig. \ref{fig:4}(a)], the links shorter than the critical length are always pinned throughout the experiment, whereas the temperature-dependent $L_{i}$ is the only relevant parameter that affects the shear modulus. That is, the dislocation network link always satisfies the pinning condition $(L_{n}<L_{c})$. On the basis of this fact, the shear modulus in the temperature scan procedure is constructed in Fig. \ref{fig:6}(a) using Eq. (\ref{eq:11}). The experimental results are displayed in Fig. \ref{fig:6}(b) for comparison.

Using the mechanical breakaway model, the temperature scan under a constant stress and the stress down-scan at a constant temperature always lead to the identical configuration and consequently the same shear modulus for a certain temperature and stress. The calculated results also seem to agree well with the expectation. However, we observed a discrepancy between the shear modulus measured via the conventional temperature scan with constant stress and the shear modulus measured via the stress down-scan below a certain characteristic temperature. The deviation provides evidence that the low-temperature shear modulus of solid helium is hysteretic. The hysteresis between these two procedures cannot be understood clearly using only the mechanical breakaway model. This issue will be discussed further in another article.\cite{22}

The vibrational motion of dislocation is hindered by various damping mechanisms, resulting in energy dissipation. The dissipated energy ratio can be expressed as the inverse of the quality factor, which is given by $Q^{-1}=\frac{1}{2\pi}\frac{\Delta W}{W}$, where $\Delta W$ is the energy lost, and $W$ is the total energy of the vibrating loop. According to Granato and Lucke,\cite{9} the energy dissipation in terms of $Q^{-1}$ is written as
\begin{equation} \label{eq:15}
Q^{-1}=R\int LQ^{-1}(L)N(L)dL.
\end{equation}
Here, $Q^{-1}(L)$ is the energy dissipation due to the dislocation loop of length $L$, given by
\begin{equation} \label{eq:16}
Q^{-1}(L)=\frac{8\mu_{0}}{\pi^{3}\rho}\frac{\omega d}{(\omega_{0}^{2}-\omega^{2})^{2} + \omega^{2}d^{2}},
\end{equation}
where $\mu_{0}$ and $\rho$ are the elastic shear modulus and the density of solid helium, respectively; $\omega_{0}$ is the resonance frequency of a vibrating dislocation loop $(=\frac{1}{L}\sqrt{\frac{2\mu_{0}}{(1-\nu)\rho}})$, $\omega$ is the frequency of the external stress, and $d$ is the damping constant. The primary damping source in solid helium at sufficiently low temperature is the dragging force induced on the dislocation lines by $^3$He impurity atoms.\cite{14} Therefore, the damping constant is proportional to the $^3$He density near the dislocation line, which is equivalent to the inverse of the pinning length,
\begin{equation} \label{eq:17}
d(T)=d_{0} L_{i}^{-1},
\end{equation}
where $d_{0}$ is a fitting parameter, which has been determined to be $2.43 \times 10^{5}$. Note that this impurity damping occurs only if the dislocation loop interacts with impurity atoms near the pinning points. When a network link is not pinned and vibrates freely, in contrast, the accompanying energy loss would be negligible. On the basis of this argument, one may calculate the energy dissipation by integrating over all the lengths of pinned loops, i.e., the lengths satisfying the condition $(L_{i}<L_{n}<L_{c})$.
\begin{equation} \label{eq:18}
Q^{-1}=R\int_{0}^{L_{c}} \int_{0}^{L_{n}} L_{i}Q^{-1}(L_{i}) N_{p}(L_{i},L_{n})dL_{i}dL_{n}.
\end{equation}
The calculated dissipation as a function of temperature is shown in Fig. \ref{fig:7}(a). As a higher external stress is applied, the fraction of pinned links, which are the source of the dissipation, is reduced, resulting in less dissipation. When an external stress is exerted on the solid helium, the resulting strain is not in phase owing to the damping but reveals some phase lag. The phase angle difference $(\Delta \phi)$ does not directly indicate the absolute values of the energy dissipation but is closely related to the dissipation for small phase differences, $Q^{-1} \sim \Delta \phi$. The measured phase angle difference between the applied stress and the measured strain is plotted in Fig. \ref{fig:7}(b) for comparison.

\section{CONCLUSION}

We proposed a microscopic description to explain the stress- and temperature-dependent shear modulus and dissipation in solid $^4$He using the mechanical breakaway model. A new effective distribution function for the pinning length involving both network node pinning and impurity pinning was constructed by superposition of network nodes on the impurity pinning loops. The shear modulus is determined by the dynamics of the pinning length distribution, which depends on the stress and temperature trajectory. The hysteresis of the shear modulus measured in the stress up/down-scan can be attributed to the path-dependent pinning length distribution. Shear modulus and dissipation data were successfully reproduced within the framework of the new description and exhibited quantitative consistency with experimental results. In addition, the effective distribution function of the pinning length can provide universal applicability to various dislocation network systems.

\begin{acknowledgments}
We gratefully acknowledge the financial support of the National Research Foundation of Korea through the Creative Research Initiatives (CSQR).
\end{acknowledgments}

\appendix
\section{The Effective Distribution}

\begin{figure}[b]
\includegraphics[width=8.5cm]{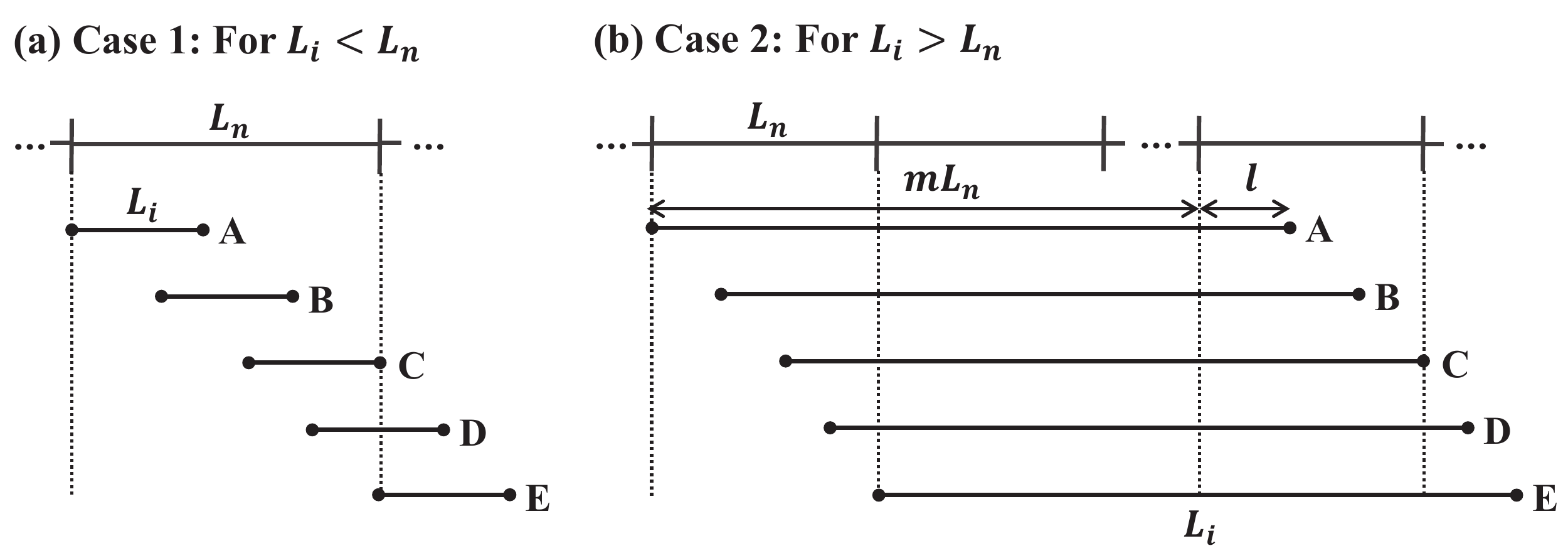}
\caption{\label{fig:A1} Schematic configurations of two cases: (a) $L_i < L_n$ and (b) $L_i > L_n$. The uppermost line represents the network line with network nodes, which is displayed as vertical line segments on the line. The horizontal line segments below represent the impurity pinning loops in various positions.}
\end{figure}

Assume a subset of dislocation network where all the network links have same length, $L_n$. Before considering the additional separation of dislocation loops by the network node, the distribution function for impurity pinning length ($L_i$) derived by Koehler can be written as
\begin{equation} \label{eq:A1}
N(L_i,L_n)dL_i = \frac{\Lambda_n}{L_{iA}^2} \exp(-\frac{L_i}{L_{iA}}) dL_i,
\end{equation}
where $\Lambda_n dL_n=L_n N_n (L_n)dL_n$ is the total length of the dislocations per unit volume and the subscript $n$ indicates that all the links have the length $L_n$, and $L_{iA}$ is the average length of $L_i$. The addition of the network node splits the impurity pinning loops into smaller loops, which changes the shape and range of the distribution function of $L_i$ significantly. The effective loop length distribution modified by the superposition of the network node can be discussed for two different cases.

\subsection{Case 1: For $L_i \leq L_n$}

For $L_i \leq L_n$, several possible configurations of the superposition are depicted in Fig.\ref{fig:A1}(a). All the configurations in the cycle (A-E) are equally probable since there is no preference for the relative position of the network node. The addition of a network node between two neighboring impurity pinning sites can divide the existing dislocation loops into two new dislocation loops as shown in the cycle (C-E). The dislocation loops remain the same otherwise. Accordingly, the probability to have the network node in the dislocation loops is given as the ratio of the impurity pinning length, $L_i$, with respect to the network link length, $L_n$.  Note that the probability can be found by considering the possibility for the position where the node is on the impurity pinning loop (C-E) divided by the possibility for any position in the whole cycle (A-E). The new distribution function can be obtained by subtracting the eliminated dislocation loops from the original distribution and including the newly created dislocation loops by the additional separation. First, the distribution function change due to the elimination, $\Delta N_- (L_i, L_n) dL_i$, is then the product of the separation probability and the distribution function for the existing impurity pinning loops, which is simply $N_i (L_i, L_n) dL_i$:
\begin{eqnarray} \label{eq:A2}
\Delta N_- (L_i, L_n) dL_i & = & -\frac{L_i}{L_n}\times N(L_i, L_n)dL_i \nonumber
\\ & = & -\frac{L_i}{L_n}\times \frac{\Lambda_n}{L_{iA}^2} \exp(-\frac{L_i}{L_{iA}}) dL_i.
\end{eqnarray}

Second, the distribution function change by the creation of the split dislocation loops can be denoted as $\Delta N_+ (L_i, L_n) L_i$. Assume an impurity pinning loop of length $L_i'$ is split into two new impurity pinning loops, one of which has the length of $L_i$ (apparently shorter than $L_i'$). The distribution function for the loops of length $L_i'$ can be obtained with the same way described above (i.e. the product of the separation probability and the distribution function for the existing impurity pinning loops): $\frac{L_i'}{L_n}\times N(L_i',L_n)dL_i'$. Since there is no preference for the length of resulting two new loops, the new loops can have any length smaller than $L_i'$. The probability that the split loop length is found in the range of $(L_i, L_i+dL_i)$ is simply given by $dL_i /L_i'$. Hence the distribution function change due to the creation can be written  by
\begin{equation} \label{eq:A3}
\begin{split}
\Delta N_{+,1} (L_i, L_n) dL_i = \int_{L_i}^{L_n} 2\times \frac{dL_i}{L_i'} \times \frac{L_i'}{L_n} \times N(L_i', L_n)dL_i'
\\ = \frac{2\Lambda_n}{L_n L_{iA}}\Big[ \exp(-\frac{L_i}{L_{iA}})-\exp(-\frac{L_n}{L_{iA}}) \Big]dL_i,
\end{split}
\end{equation}
where the factor 2 indicates that always two loops are created by the superposition. Note that the range of $L_i'$ is chosen to be $(L_i, L_n)$ since only impurity pinning loops longer than $L_i$ can contribute to the creation of new loops of length $L_i$. 

\subsection{Case 2: For $L_i > L_n$}

For the impurity pinning loops longer than $L_n$, the situation becomes slightly different. All the impurity pinning loops should be split by the superposed nodes at least once and the resulting loops  have length less than or equal to $L_n$. Assume an impurity pinning loop of length $L_i' (> L_n)$. The length $L_i'$ can be written in the following form:
\begin{equation} \label{eq:A4}
L_i' = mL_n + l, \;\; \; 0\leq l<L_n, \;\; m=1,2,3,\cdots .
\end{equation}
where $m$ is the number of the resulting loops of length $L_n$ and $l$ is the loop length of the remainder.

First, note that two new loops shorter than $L_n$ are created at both ends of the impurity pinned dislocation loop of length $L_i'$ (see Fig. \ref{fig:A1}(b)). The length of the two split dislocation loops at the both ends varies from zero to $L_n$ in the (A-E) cycle of which value is equally probable between zero and $L_n$. The probability that the new loop has a length in the range of $(L_i, L_i+dL_i)$ is given as $dL_i/L_n$. The distribution function change due to the creation of loops at both ends is then
\begin{eqnarray} \label{eq:A5}
\Delta N_{+,2} (L_i, L_n) dL_i &=& \int_{L_n}^{\infty} 2\times \frac{dL_i}{L_n} \times N(L_i', L_n)dL_i' \nonumber
\\ & = & \frac{2\Lambda_n}{L_n L_{iA}}\exp(-\frac{L_n}{L_{iA}})dL_i,
\end{eqnarray}
where the factor 2 indicates that two loops are created from the loop of length $L_i'$. Note that the range of $L_i'$ is chosen to be $(L_n, \infty)$ since only the loops longer than $L_n$ are taken into account here.

 Adding all the distribution function changes to the original distribution function, the new distribution function is obtained as 
\begin{eqnarray} \label{eq:A6}
N_p(L_i,L_n)dL_i & = & N(L_i,L_n)dL_i + \Delta N_- (L_i,L_n)dL_i \nonumber
\\ & & + \Delta N_{+,1}(L_i,L_n)dL_i+ \Delta N_{+,2} (L_i,L_n)dL_i \nonumber
\\ & = & (1-\frac{L_i}{L_n}+\frac{2L_{iA}}{L_n})\frac{\Lambda_n}{L_{iA}^2}\exp(-\frac{L_i}{L_{iA}})dL_i.
\end{eqnarray}
Noteworthy is that the new distribution function does not have the dislocation loops longer than $L_n$, since the pinning length cannot be greater than the network pinning length. 

However, the dislocation links pinned only with the network nodes are not included in the new distribution function in the Eq. (\ref{eq:A6}). The additional creation of the loops with $L_n$ for an impurity pinning loop of length $L_i'=mL_n+l$ results in either $(m-1)$ loops of length $L_n$ (A-C in Fig. \ref{fig:A1}(b)) or $m$ loops of length $L_n$ (C-E). Since the position of nodes can be in any configurations in the cycle (A-E), the probability that $L_i'$ is divided into $m$ loops of length $L_n$ is simply the number of possible configurations for (C-E) divided by that for (A-E), which is given by $\frac{l}{L_n} = \frac{L_i'-mL_n}{L_n}$. The probability for $(m-1)$ loops with length $L_n$ is then $(1-\frac{l}{L_n})=\frac{(m+1) L_n-L_i'}{L_n}$. Now, the number of resulting loops of length $L_n$ created by the superposition of nodes on the loop of a length in the range of $(L_i', L_i' + dL_i')$ is calculated by the product of the expectation number of new loops and the distribution function of the existing loops.
\begin{widetext}
\begin{eqnarray} \label{eq:A7}
N_{L_n}(L_i',L_n)dL_i'& = & \Big[ m\times \frac{L_i'-mL_n}{L_n} + (m-1)\times \frac{(m+1)L_n - L_i'}{L_n} \Big] N(L_i',L_n)dL_i' \nonumber
\\ & = & (\frac{L_i'}{L_n}-1)\frac{\Lambda_n}{L_{iA}^2}\exp(-\frac{L_i'}{L_{iA}})dL_i',
\end{eqnarray}
\end{widetext}
Integration of the Eq. (\ref{eq:A7}) over the range of $L_i' (L_n < L_i' < \infty)$ will lead to the total number of resulting loops of length $L_n$.

\begin{equation} \label{eq:A8}
N_u(L_n) = \int_{L_n}^{\infty} N_{L_n}(L_i',L_n)dL_i' = \frac{\Lambda_n}{L_n}\exp(-\frac{L_n}{L_{iA}})
\end{equation}
In conclusion, the new effective distribution function is written as below.
\begin{widetext}
\begin{eqnarray} \label{eq:A9}
N_p(L_i,L_n) & = & (1-\frac{L_i}{L_n}+\frac{2L_{iA}}{L_n})\frac{\Lambda_n}{L_{iA}^2} \exp(-\frac{L_i}{L_{iA}})dL_i, \textrm{for } 0<L_i<L_n \nonumber
\\ N_u(L_n) & = & \frac{\Lambda_n}{L_n}\exp(-\frac{L_n}{L_{iA}})
\end{eqnarray}
\end{widetext}
The validity of the effective distribution function can be readily tested with the normalization condition.

\begin{equation} \label{eq:A10}
\int_{0}^{L_n} L_i N_p(L_i,L_n)dL_i + L_n N_u(L_n) = \Lambda_n
\end{equation}
The first term on the left hand side of the Eq. (\ref{eq:A10}) is the total length of new loops whose length is less than $L_n$ and the second term is the total length of new loops of length $L_n$.

\section{Characteristics of the Dislocation Network}

The new effective distribution function (\ref{eq:A10}) is composed of an impurity-pinned part $(L_i < L_n)$ and naturally unpinned part $(L_i = L_n)$ that is only pinned with the network nodes. In the following paragraph, the characteristics of each part will be discussed to help understanding the article.

\subsection{Impurity-pinned part $(L_i < L_n)$}

The total number of the loops shorter than $L_n$ with the effective distribution function can be obtained by integrating the distribution function of the impurity-pinned part.

\begin{equation} \label{eq:B1}
\int_{0}^{L_n} N_p(L_i,L_n)dL_i
= \Lambda_n [ \frac{1}{L_{iA}}+\frac{1}{L_n}-\frac{1}{L_n} \exp(-\frac{L_n}{L_{iA}}) ]
\end{equation}

The total length of the loops shorter than $L_n$ with the effective distribution function can be given by integrating the product of the length and the distribution function.

\begin{equation} \label{eq:B2}
\int_{0}^{L_n} L_i N_p(L_i,L_n)dL_i
= \Lambda_n [ 1-\exp(-\frac{L_n}{L_{iA}}) ]
\end{equation}

The total number of the impurity-pinned network links is then given by the total length of the loops shorter than $L_n$ divided by $L_n$.

\begin{equation} \label{eq:B3}
\frac{\Lambda_n}{L_n} [ 1-\exp(-\frac{L_n}{L_{iA}}) ]
\end{equation}

The average number of the impurity pinning loops in a pinned network link may be calculated by dividing the total number of the impurity pinning loops (B1) by the number of the pinned links (B3).

\begin{equation} \label{eq:B4}
\frac{\Lambda_n[\frac{1}{L_{iA}}+\frac{1}{L_n}-\frac{1}{L_n} \exp(-\frac{L_n}{L_{iA}})]}{\frac{\Lambda_n}{L_n} [ 1-\exp(-\frac{L_n}{L_{iA}})]} = \frac {L_n}{L_{iA}[ 1-\exp(-\frac{L_n}{L_{iA}})]}+1
\end{equation}

Since the number of the impurity pinning points in a network link is one less than the number of the impurity pinning loops, the number of the pinning points $k$ is therefore

\begin{equation} \label{eq:B5}
k = \frac {L_n}{L_{iA}[ 1-\exp(-\frac{L_n}{L_{iA}})]}.
\end{equation}

\subsection{Unpinned part $(L_i = L_n)$}

The total number of the dislocation links of length $L_n$ with the effective distribution function is

\begin{equation} \label{eq:B6}
N_u(L_n) = \frac {\Lambda_n}{L_n}\exp(-\frac{L_n}{L_{iA}}).
\end{equation}

The total length of the links of length $L_n$ with the effective distribution function is then obtained simply by multiplying $L_n$.

\begin{equation} \label{eq:B7}
L_n N_u(L_n) = \Lambda_n\exp(-\frac{L_n}{L_{iA}})
\end{equation}

Now one can notice that the total number of the all kinds of pinning loops is obtained by adding (B1) and (B6).

\begin{equation} \label{eq:B8}
\Lambda_n(\frac{1}{L_{iA}}+\frac{1}{L_n})
\end{equation}

Moreover, the average length of the pinning loops is then

\begin{equation} \label{eq:B9}
\frac{L_{iA} L_n}{L_{iA}+L_n}.
\end{equation}

Note that the averaged results, Eq. (\ref{eq:B8}) and Eq. (\ref{eq:B9}), are the same as those obtained in the dislocation-vibration model.\cite{14}

\bibliography{MicroD}

\end{document}